\documentclass[twocolumn,prl,showpacs,aps]{revtex4-1}
\usepackage{amsmath,amsfonts,braket,graphicx,array}
\newcolumntype{P}[1]{>{\centering\arraybackslash}p{#1}}
\begin{document}
\title{\bf Oscillating nuclear electric dipole moment induced by axion dark matter produces atomic and molecular  EDM}
\author{V. V. Flambaum
$^{1,2,3}$
} 
\affiliation{
$^1$
School of Physics, University of New South Wales,  Sydney 2052,  Australia}
\affiliation{$^2$Helmholtz Institute Mainz, Johannes Gutenberg-Universit\"at Mainz, 55099 Mainz, Germany}
\affiliation{$^3$The New Zealand Institute for Advanced Study, Massey University Auckland, 0632 Auckland, New Zealand}
\author{H. B. Tran Tan}
\affiliation{School of Physics, University of New South Wales, Sydney 2052, Australia}
\date{\today}

\begin{abstract}
According to the Schiff theorem, a nuclear electric dipole moment (EDM) is completely shielded in a neutral atom by electrons. We consider the extension of Schiff theorem to the cases of time-dependent external electric fields and nuclear EDMs. A time-dependent external electric field penetrates to the nucleus and causes nuclear spin rotation. 
 Interaction with the axion dark matter field generates nuclear EDM $d=d_0 \cos (\omega t)$ oscillating with the frequency $\omega= m_a c^2/\hbar$. This EDM generates atomic and molecular EDM proportional to $\omega^2$. However, this EDM does not lead to the nuclear spin rotation in the constant external electric field. Nevertheless, if the nuclear EDM and the external electric field oscillate with the same frequency then the nuclear spin rotation angle grows linearly with time. The molecular EDMs induced by nuclear EDMs are strongly enhanced since nuclei move slowly and do not produce as efficient screening of oscillating nuclear EDM as electrons do. An additional strong enhancement comes from the small energy interval between rotational molecular levels. Finally, if the nuclear EDM oscillation frequency is in resonance with a molecular transition, there may be a significant resonance enhancement. Numerical estimates for the molecules HF, LiF, YbF, BaF, TlF, HfF${}^+$, ThF${}^+$, ThO and WC are provided.
 \end{abstract}

\maketitle

\textbf{Introduction:} It was suggested in Ref.\ \cite{Graham} that interaction with the axionic dark matter produces oscillating neutron and nuclear electric dipole moments. However, according to the Schiff theorem  \cite{Schiff}, the nuclear EDM is completely screened in neutral atomic systems. Atomic and molecular EDMs are actually produced by the nuclear Schiff moment which is suppressed compared to EDM by an additional second power of the nuclear radius which is very small on the atomic scale  \cite{Sandars,Hinds,SFK,FKS1985,FKS1986} (see also \cite{Khriplovich,Auerbach,FGP,FlambaumKozlov,Sandars1965,Flambaum,SushkovFlambaum} for other effects producing atomic and molecular EDM). The effects produced by the axion-induced Schiff moment have been considered in Ref. \cite{Stadnik2014}.  A corresponding experiment in solids has been proposed in  Ref.\ \cite{Casper}. The first results of the oscillating neutron EDM and Hg atom's EDM measurements are presented in Ref.\ \cite{nEDM} where the limits on the low-mass axion interaction constant with matter have been improved up to three orders of magnitude. 

 In the present paper it is shown that an oscillating nuclear EDM such as that produced by the axion dark matter is not completely screened in atoms and molecules and produces atomic and molecular EDMs. The latter case is especially interesting since the effect in molecules is several orders of magnitude larger than in atoms. Indeed, in the screening of the static nuclear EDM, the nuclei in a molecule play as important a role as the electrons. If the nuclear EDM oscillates, because nuclei are not as fast-moving as the electrons, the screening is incomplete. As a result, the residual, partly screened EDM in molecules is $M_N/m_e$ times larger than that in atoms. Here $M_N$ is the nuclear mass and $m_e$ is the electron mass. Enhancement of the oscillating nuclear EDM may happen if the oscillation frequency is in resonance with a molecular transition frequency. 
 
\textbf{Screening theorem for time-dependent electric field and EDM:}
As known, a nucleus in a neutral system (atom or molecule) is completely screened from a constant electric field\ \cite{Schiff}. We will here present a derivation of this fact following the Appendix in Ref.\ \cite{Spevak1997}. For definiteness, we assume that the system in question is a neutral atom in a static homogeneous external electric field of an arbitrary strength (we ignore the possibility of atomic ionization and effects of magnetic fields).

The Hamiltonian of an atom placed in a static homogeneous external electric field ${\bf E}_0$ is
\begin{equation}
\begin{aligned}
H&=\sum_i\left[K_i-e\phi_0\left({\bf r}_i\right)+e{\bf r}_i\cdot{\bf E}_0\right]\\
&+\sum_{i>j}\frac{e^2}{\left|{\bf r}_i-{\bf r}_j\right|}-{\bf d}\cdot{\bf E}_0,
\end{aligned}
\end{equation}
where $K_i$ and ${\bf r}_i$ are the kinetic energy and coordinates of the electrons, ${\bf d}$ is the static nuclear EDM and $\phi_0\left({\bf r}_i\right)$ is the electrostatic nuclear potential given by
\begin{equation}
\phi_0\left({\bf r}_i\right)=e\int\frac{\rho\left({\bf r}\right)d^3{\bf r}}{\left|{\bf r}_i-{\bf r}\right|},
\end{equation}
where $\rho$ is the nuclear charge distribution. We consider here the case of an infinitely heavy nucleus. The nuclear recoil correction is not enough to generate an atomic EDM\ \cite{Schiff}.

We add to $H$ and auxiliary term 
\begin{equation}\label{AuxiliaryV}
V={\bf d}\cdot{\bf E}_0-\frac{1}{Ze}\sum_i{\bf d}\cdot\nabla_i\phi_0\left({\bf r}_i\right)\,,
\end{equation}
which, in the linear approximation in $\bf d$, does not produce any energy shift, $\left<V\right>=0$. Indeed, we have
\begin{equation}
\frac{i}{m}\left[\sum_i{\bf p}_i,H\right]=-e\sum_i\nabla_i\phi_0\left({\bf r}_i\right)+Ze{\bf E}_0,
\end{equation}
where we have taken into account the fact that the total electron momentum $\sum_i{\bf p}_i$ commutes with the electron-electron interaction term. Using Eq.\ \eqref{AuxiliaryV} and the fact that 
\begin{equation}\label{Commutator}
\frac{i}{m}\bra{\psi}\left[\sum_i{\bf p}_i,H\right]\ket{\psi}\propto\left(E_{\psi}-E_{\psi}\right)=0
\end{equation}
($\psi$ is the wavefunction of the Hamiltonian $H$), we obtain
\begin{equation}
\left<V\right>=\left<{\bf d}\cdot{\bf E}_0-\frac{1}{Ze}\sum_i{\bf d}\cdot\nabla_i\phi_0\left({\bf r}_i\right)\right>=0.
\end{equation}

To find an EDM one needs to measure a linear energy shift in an external electric field. Since V does not contribute to this shift we can add it to the Hamiltonian
\begin{equation}
\begin{aligned}
\tilde{H}&\equiv H+V\\
&=\sum_i\left[K_i-e\phi\left({\bf r}_i\right)+e{\bf r}_i\cdot{\bf E}_0\right]+\sum_{i>j}\frac{e^2}{\left|{\bf r}_i-{\bf r}_j\right|}\,,
\end{aligned}
\end{equation}
where
\begin{equation}
\phi\left({\bf r}_i\right)=\phi_0\left({\bf r}_i\right)+\frac{1}{Ze}{\bf d}\cdot\nabla_i\phi_0\left({\bf r}_i\right)\,.
\end{equation}

Note that the Hamiltonian $\tilde{H}$ does not contain the direct interaction ${\bf d}\cdot{\bf E}_0$ between the nuclear EDM and external field (Schiff theorem). The dipole term is also canceled out in the multipole expansion of $\phi\left({\bf r}_i\right)$.

Let us now consider the case where the nuclear EDM is time-dependent ${\bf d}={\bf d}\left(t\right)$. In this case, Eq.\ \eqref{Commutator} becomes
\begin{equation}
\begin{aligned}
\frac{i}{m}\left<\left[\sum_i{\bf p}_i,H\right]\right>&=-\frac{1}{m}\frac{d}{dt}\left<\sum_i{\bf p}_i\right>\\
&=\frac{1}{m}\frac{d\left<{\bf p}_{\rm nuc}\right>}{dt}\propto {\bf d}\,.
\end{aligned}
\end{equation}
Therefore, the contribution due to $V$ is zero in the first order in $\bf d$. As a result, just as in the case of a static nuclear EDM, there is no direct interaction between a time-dependent nuclear EDM and a static external electric field, hence, no nuclear spin rotation. Indeed, the external electric field does not penetrate to the nucleus (since an atom and its nucleus are not accelerated by a static homogeneous electric field), so the nuclear EDM has nothing to interact with.

Now consider the case of a time-dependent electric field. In this case, we have
\begin{equation}
\frac{1}{m}\frac{d\left<{\bf p}_{\rm nuc}\right>}{dt}\propto {\bf E}_0\,,
\end{equation} since the external field now penetrates to the nucleus\ \cite{Flambaum2018,FlambaumSamsonov2018,TranFlambaum2019}. Indeed, the external electric field forces the electron shells to oscillate and since the atom's center of mass stays at rest, the nucleus must move, so the electric field on it is not zero. Therefore, the nuclear EDM interacts with this electric field and nuclear spin rotation happens.

Note that the absence of nuclear spin rotation in the case of a static electric field does not mean that the oscillating nuclear EDM does not produce any effect. An oscillating nuclear EDM excites the electrons and produces atomic and molecular EDMs (as demonstrated below). This effect is particularly clear in the case where the nuclear EDM's frequency of oscillation is in resonance with some atomic or molecular frequency, in which case the electronic wavefunciton is a linear combination of two states of opposite parities and thus gives rise to oscillating atomic and molecular EDMs. Oscilalting nuclear EDMs may be detected using the atomic and molecular transitions they induced, as investigated in Ref.\ \cite{Wickenbrock2019,FlambaumTransition2019}.

The case where both the nuclear EDM and the external electric field are time-dependent, particularly when they are oscillating, is of special interest. As demonstrated in Refs.\ \cite{Flambaum2018,FlambaumSamsonov2018,TranFlambaum2019}, an external electric field which oscillates with a frequency $\omega$, ${\bf E}_0\sim \cos\omega t$, induces an electric field on the nucleus which oscillates with the same frequency. The interaction of this field with a nuclear EDM which itself oscillates with a frequency $\Omega$, ${\bf d}\sim\cos\Omega t$, is proportional to $\cos\omega t\cos\Omega t$. If $\omega=\Omega$ then this interaction contains a time-independent component and the nuclear spin rotation angle grows linearly with time.
  
\textbf{Nuclear EDM produced by the axion dark matter field:} It has been noted in Ref. \cite{Witten} that the neutron EDM may be produced by the QCD $\theta$ term. Numerous references  and recent results for the neutron and proton EDMs are summarised in Ref.\ \cite{Yamanaka}:
 \begin{equation}\label{theta}
 \begin{aligned}
 d_n&= -(2.7 \pm 1.2) \times 10^{-16} \theta \, e\,  \textrm{cm} \,,\\
 d_p&= (2.1 \pm 1.2) \times 10^{-16} \theta \, e\,  \textrm{cm}\,.
 \end{aligned}
\end{equation}

Calculations of the nuclear EDM produced by the P,T-odd nuclear forces have been performed in the Refs.\ \cite{HH,SFK,FKS1985,FKS1986}. For a general estimate of the nuclear EDM it is convenient to use a single-valence-nucleon formula from Ref.\ \cite{SFK} and express the result in terms of $\theta$ following Ref.\ \cite{FDK}:
\begin{equation}\label{d}
   d \approx   e\left(q- \frac{Z}{A}\right)\left(1- 2q\right)  \xi \left<\sigma\right> \,,
  \end{equation}
  where $\xi = 7 \times 10^{-16}  \theta{\rm cm}$.  
  
  Here $q=1$ for the valence proton, $q=0$ for the valence neutron, the nuclear spin matrix element $\left<\sigma\right>=1$ if $I=l+1/2$ and $\left<\sigma\right>=-I/\left(I+1\right)$ if $I=l-1/2$. Here, $I$ and $l$ are the total and orbital momenta of the valence nucleon.

It was noted in Ref.\ \cite{Graham} that the axion dark matter field may be an oscillating $\theta$ term and thus generates the oscillating neutron EDM. To reproduce the density of dark matter, following Ref.\ \cite{Stadnik2014} we may substitute $\theta(t)=\theta _0 \cos(\omega t)$ where $\theta _0= 4 \times 10^{-18}$, $\omega=m_a c^2/\hbar$ and $m_a$ is the axion mass. In the following sections, we estimate the electric dipole moment of atoms and molecules induced by the oscillating nuclear EDM.

\textbf{Atomic EDM  induced by an oscillating nuclear EDM:} The Hamiltonian of an atom in the field of an oscillating nuclear EDM ${\bf d}={\bf d}_0 \cos (\omega t) $ may be written as 
\begin{equation}\label{Phi}
V= 
e\sum_{k=1}^{N_e}\frac{{\bf d\cdot r}_k}{r_k^3}=\frac{i}{Z e \hbar}[{\bf P\cdot d},H_0] \,,
\end{equation}
where $H_0$ is the Schr{\"o}dinger or the Dirac Hamiltonian for the atomic electrons in the absence of $\bf d$, $N_e$ is the number of electrons, $Ze$ is the nuclear charge, $Z_i=Z-N$,  $-e$ is the electron charge, $r_k$ is  the electron position relative to the nucleus, ${\bf P}=\sum_{k=1}^{N_e}{\bf p_{k}}$ is the total momentum of all atomic electrons (which commutes with the electron-electron interaction but not with the nuclear-elect interaction $U=-\sum_{k=1}^{N_e}Ze^2/r_k$: $[{\bf P },H_0] = [{\bf P }, U] = -i \hbar  Z e^2 \sum_{k=1}^{N_e}\nabla \frac{1}{r_k}$). Here we assumed that the nuclear mass is infinite and neglect very small effects of the  Breit and  magnetic interactions. 

Using  $H_0 \ket{n} = E_n\ket{n}$  we obtain the matrix element of $V$ between atomic states $\ket{n}$ and $\ket{m}$
\begin{equation}\label{V12}
\bra{n}V\ket{m}=\frac{iE_{nm}}{Ze \hbar}\bra{n}{\bf P\cdot d}\ket{m}\,,
\end{equation}
where $E_{nm}=E_n-E_m$. 

Using the time dependent perturbation theory\ \cite{Landau} for the oscillating  perturbation $V=V_0 \cos{ \omega t}$ and Eq.\ \eqref{V12} we obtain a formula for the induced atomic EDM
\begin{equation}\label{HE}
\begin{aligned}
{\bf D}_{\rm ind}&=  2\sum_n \frac{E_{0n}{\rm Re}\left(\bra{0}V\ket{n}\bra{n}{\bf D}\ket{0}\right)}{E_{0n}^2 -\epsilon^2}\\
 &=\frac{2}{Z e\hbar}\sum_n \frac{E_{0n}^2{\rm Im}\left(\bra{0}{\bf P\cdot d}\ket{n}\bra{n}{\bf D}\ket{0}\right)}{E_{0n}^2 -\epsilon^2}\,,
\end{aligned}
\end{equation}
where $\epsilon =\hbar \omega$ and ${\bf D}=-e \sum_{k=1}^{N} {\bf r_k}$. 

The energy dependent factor  may be presented as
\begin{equation}\label{energy}
\frac{E_{0n}^2}{E_{0n}^2 -\epsilon^2}=1 + \frac{\epsilon^2}{E_{0n}^2 -\epsilon^2}\,.
\end{equation}
The energy independent term 1 on the right hand side allows us to sum over states $\ket{n}$ in Eq.\ \eqref{HE}. Using the closure condition and the commutator relation $[{\bf P},{\bf D}]= -i e \hbar N_e$, this term gives
\begin{equation}\label{shielding}
\begin{aligned}
{\bf D}_{\rm atom}&={\bf d} +{\bf D}_{\rm ind}=\frac{Z_i}{Z}{\bf d}\\
  & + 
 \frac{2}{Z e\hbar}\sum_n \frac{\epsilon^2{\rm Im}\left(\bra{0}{\bf P\cdot d}\ket{n}\bra{n}{\bf D}\ket{0}\right)}{E_{0n}^2 -\epsilon^2}\,.\\
 \end{aligned}
\end{equation}

We observe that, in agreement with the Schiff theorem, the atomic electric dipole moment ${\bf D}_{\rm atom}$ vanishes in a neutral atom ($Z_i=Z-N=0$) with static nuclear EDM ($\epsilon=\hbar \omega=0$).
 
 Assume that nuclear EDM $d$ is directed along the $z$-axis.
  Using the non-relativistic commutator relation  ${\bf P}=- \frac{i m_e} {e \hbar} [H_0,{\bf D}]$ (where $m_e$ is the electron mass), we can express the atomic EDM in terms of the atomic dynamical polarisability $\alpha_{zz}(\omega)$
\begin{equation}\label{polarizability}
\begin{aligned}
D^z_{\rm atom}&=  \frac{d_z}{Z}\left(Z_i -
 \frac{m_e\epsilon^2 \alpha_{zz}}{e^2\hbar^2}\right)\,,\\
 \alpha_{zz}&=2 \sum_n \frac{E_{n0}\left|\bra{0}D_z\ket{n}\right|^2}{E_{n0}^2 -\epsilon^2}\,.
 \end{aligned}
\end{equation}

The axion field oscillation frequency may be very small on the atomic scale, therefore, we may use static polarisabilities in this expression which are known for all atoms. 
 The formula\ \eqref{polarizability} may be rewritten, with the energy and the polarizabilty expressed in atomic units ${\tilde \epsilon} = \frac{\epsilon}{e^2/a_B}$ and ${\tilde  \alpha}_{zz}=\frac{ \alpha_{zz}}{a_B^3}$ (where $a_B$ is the Bohr radius), as:  
 \begin{equation}\label{polarizabilityAtomic}
D^z_{\rm atom}= 
\frac{Z_i - {\tilde \epsilon}^2 {\bf \tilde  \alpha}_{zz}}{Z}d_z
\end{equation}

Since the atomic EDM  $D_{\rm atom}$  is proportional to $1/Z$, it appears that the shielding is stronger in heavy atoms. This, however, is not necessary the case since, for example in hydrogen and helium  ${\bf \tilde  \alpha}_{zz} \sim 1$ whereas ${\bf \tilde  \alpha}_{zz} \sim 400$ in caesium ($Z$=55).  Indeed, the numerical value  of the polarizability ${\tilde  \alpha}_{zz}$ in atomic units often exceeds the value of the nuclear charge $Z$, therefore, the suppression of EDM in a neutral atom mainly comes from the small frequency of the dark matter field oscillations in atomic units,  ${\tilde \epsilon}$.

\textbf{Molecular EDM induced by oscillating nuclear EDM:} We see from the first line in Eq. (\ref{polarizability}) that the residual EDM in a neutral system $Z_i=0$ is proportional to the mass $m$ of the particle which produces the screening of the nuclear EDM $\bf d$. Masses of nuclei $M_N$ in a molecule are up to 5 orders of magnitude larger than the mass of electron $m_e$. In addition, the interval between molecular rotational energy levels ($\sim m_e/M_N$ atomic units) are many orders of magnitude smaller than typical energy intervals in atoms and this may give  an additional enormous advantage, see the denominator in the second line in Eq. (\ref{polarizability}). Finally, since the molecular spectra are very rich, the energy intervals are small and may be tuned by electric and magnetic fields, it is easier to bring them into resonance with the small oscillation frequency of the axion dark matter field.

Calculations presented in Appendix A give the following results for the induced electric dipole of a neutral diatomic molecule when $\epsilon$ is smaller or of the order of the first rotational energy $E_{\rm rot}$
\begin{equation}\label{D_mol_EDM}
\mathbf{D}_{\text{mol}}^{\text{EDM}}\approx \frac{2{{\mu }_{N}}\bar{X}\bar{d}E_{\text{rot}}}{3{{e}}{{\hbar }^{2}}}\frac{\epsilon ^{2}}{E_{\text{rot}}^{2}-{{\epsilon }^{2}}}\left( \frac{{{\mathbf{d}}_{1}}}{{{Z}_{1}}}-\frac{{{\mathbf{d}}_{2}}}{{{Z}_{2}}} \right)\,,
\end{equation}
where $\mu_N=M_1M_2/\left(M_1+M_2\right)$ is the reduced nuclear mass, $\bar{X}$ is the ground state inter-nuclear distance, $\bar{d}$ is the ground state intrinsic electric dipole of a polar molecule and $E_{\rm rot} \approx \hbar^2\mu_N^{-1}\bar{X}^{-2}$ is the energy of the first rotational state and ${\bf d}_{1,2}$ are the nuclear EDMs. In writing Eq.\ \eqref{D_mol_EDM}, we have assumed that the molecular ground state has total angular momentum 0. 

Note that traditionally, the interaction of the nuclear EDMs and a molecule is expressed in terms of the nuclear spin-molecular axis interaction. To do this, we need to rewrite Eq.\ \eqref{D_mol_EDM} in terms of the polarization degree of the molecule in an electric field $\boldsymbol{\mathcal{E}}$, ${\bf P}=\bar{d}\boldsymbol{\mathcal{E}}/(3E_{\rm rot})$, and the energy shift $\Delta E ={\bf D}^{\rm EDM}_{\rm mol}\boldsymbol{\mathcal{E}}$. Substituting these quantities into Eq.\ \eqref{D_mol_EDM}, we have
\begin{equation}\label{D_mol_EDM_new}
\Delta E\approx \frac{2{{\mu }_{N}}\bar{X}E^2_{\text{rot}}}{e{{\hbar }^{2}}}\frac{\epsilon ^{2}}{E_{\text{rot}}^{2}-{{\epsilon }^{2}}}\left( \frac{{{\mathbf{P\cdot d}}_{1}}}{{{Z}_{1}}}-\frac{{{\mathbf{P\cdot d}}_{2}}}{{{Z}_{2}}} \right)\,.
\end{equation}

For $d_1\sim d_2$, we see that the lighter nucleus gives dominating contribution. In other words, if $Z_1\ll Z_2$ then the term ${\bf d}_2/Z_2$ drops out. We assume this is the case. In the limits $\epsilon\ll E_{\rm rot}$ and $\epsilon\gg E_{\rm rot}$, Eq.\ \eqref{D_mol_EDM} gives
\begin{equation}\label{limits}
\frac{\mathbf{D}_{\text{mol}}^{\text{EDM}}}{{{\mathbf{d}}_{1}}}\approx \left\{ \begin{matrix}
   \frac{2{{\epsilon }^{2}}\mu _{N}^{2}{{{\bar{X}}}^{3}}\bar{d}}{3{{e}}{{\hbar }^{4}}{{Z}_{1}}} & \epsilon \ll {{E}_{\text{rot}}}  \,,\\
   \frac{2\bar{d}}{3{e}{{Z}_{1}}\bar{X}} & \epsilon \gg {{E}_{\text{rot}}}  \,.\\
\end{matrix} \right.
\end{equation}

We see that in the small axion mass limit ($\epsilon=m_ac^2 \ll E_{\rm rot}$), heavy molecules have an advantage ($\mu_N^2/Z_1$). In the large axion mass limit ($\epsilon=m_ac^2 \gg E_{\rm rot}$), the ratio of the EDMs is independent of $\epsilon$ and has asymptotic value $2\bar{d}/\left(3eZ_1\bar{X}\right)< 2/\left(3Z_1\right) \leq 2/3$ ($\bar{d}\sim e\bar{X}$ for polar molecule) so molecules with at least one light nucleus are more advantageous.

The result\ \eqref{D_mol_EDM} applies for the off-resonance case. If $\epsilon = E_{\rm rot}$ then we have the following relation between the oscillation amplitudes of $\mathbf{D}_{\text{mol}}^{\text{EDM}}$ and $\mathbf{d}_{1}$;
\begin{equation}\label{res}
D_{\text{mol}}^{\text{EDM}}\approx \frac{2\bar{d}}{3{{e}}{{Z}_{1}}\bar{X}}\frac{{{E}_{\text{rot}}}}{{{\Gamma }}}{{d}_{1}}\,,
\end{equation} 
which is the large axion mass asymtotic value in Eq.\ \eqref{limits} multiplied by the resonace enhancement factor $E_{\rm rot}/\Gamma$ where $\Gamma$ is the width. Again, we see that molecules with at least one light nucleus give bigger effect.

There may be different contributions to $\Gamma$: natural width (which is typically small), Doppler width, collision width and time of flight (if the experiment is done with molecular beam). If, however, the experiment uses a trapped molecule then $\Gamma$ is mainly due to the velocity distribution of the axion: $\Gamma/E_{\rm rot}\approx \left<v\right>^2/c^2 \sim 10^{-6}$ where $\left<v\right>$ is the mean axion velocity. 

If molecules with Cesium or heavier nuclei are used then the contribution to the total ${\bf D}_{\rm mol}$ due to the Schiff moment may becomes significant. Still assuming that $\epsilon \lesssim E_{\rm rot}$, the contribution to ${\bf D}_{\rm mol}$ from the Schiff moment ${\bf S}= S{\bf I}/I$ ($\bf I$ is the nuclear total angular momentum) is
\begin{equation}\label{D_SCHIFF}
\mathbf{D}_{\text{mol}}^{\text{SCHIFF}}\approx \frac{2\bar{d}{{W}_{S}}\mathbf{S}}{3}\frac{{{E}_{\text{rot}}}}{E_{\text{rot}}^{2}-{{\epsilon }^{2}}}\,,
\end{equation}
where $W_S$ is the effective strength of the interaction between $\bf S$ and the molecular axis. We note that since $W_S$ scales as $Z^n$ with $n>2$\ \cite{SFK}, the contribution due to the heavier nucleus dominates: ${\bf S} \approx {\bf S}_2$.

To compare the effects of the nuclear EDMs and nuclear Schiff moments, it is convenient to form the ratio
\begin{equation}\label{ratio}
\frac{\mathbf{D}_{\text{mol}}^{\text{EDM}}}{\mathbf{D}_{\text{mol}}^{\text{SCHIFF}}}=\frac{{{\epsilon }^{2}}{{\mu }_{N}}\bar{X}{{d}_{1}}}{e{{\hbar }^{2}}{{Z}_{1}}{{W}_{S}}{{S}_{2}}}\,.
\end{equation}
We see that the effect of the nuclear EDMs dominates for large axion mass. Also, as noted above, for light nuclei, $W_S$ is typically small so the effect of the nuclear Schiff moment is negligible compared to that of the nuclear EDM.

In order to estimate the ratio $d_1/S_2$, we need in addition to Eq.\ \eqref{d} for the nuclear EDM, a formula for the nuclear Schiff moment $S$, which, in the case of a spherical nucleus with one unpaired nucleon, reads\ \cite{SFK}
\begin{equation}\label{Schiff}
S=-\frac{eq}{10}\xi\left[\left(t_I+\frac{1}{I+1}\right)\langle r^2\rangle-\frac{5}{3}t_Ir_q^2\right]\,,
\end{equation}
where $q$, $\xi$ and $t_I$ are defined as in Eq.\ \eqref{d}, $\langle r^2\rangle$ is the mean squared radius of the unpaired nucleon and $r^2_q$ is the mean squared charge radius. Approximately, $\langle r^2\rangle\approx r_q^2\approx(3/5)R^2$ where $R$ is the mean radius of the nucleus.


As examples, we consider the molecules H${}^1$F${}^{19}$, Li${}^7$F${}^{19}$, Yb${}^{174,176}$F${}^{19}$, Ba${}^{132,134,136,138}$F${}^{19}$, Tl${}^{203,205}$F${}^{19}$, Hf${}^{180}$F${}^{19+}$, Th${}^{232}$F${}^{19+}$, Th${}^{232}$O${}^{17}$ and W${}^{184,186}$C${}^{13}$. In LiF, the effect of the Schiff moment comes from the fluorine nucleus which is the heavier of the two whereas in TlF it comes from the thallium nucleus. We demonstrate below that ${\bf D}^{\rm LiF}_{\rm mol}$ dominates over ${\bf D}^{\rm SCHIFF}_{\rm LiF}$ for the axion mass $\epsilon \sim 10^{-5}-10^{-3}\,{\rm eV}$ whereas, due to the large Schiff moment of Tl, ${\bf D}^{\rm SCHIFF}_{\rm TlF}$ dominates over ${\bf D}^{\rm EDM}_{\rm TlF}$ for $\epsilon \lesssim 10^{-4}\,{\rm eV}$. In the other molecules, the heavier nuclei have zero spin so the Schiff moment contribution comes from the lighter nuclei (F, O and C). As a result, just as in the case of LiF, the Schiff moment contribution in these molecules is negligible in comparison with the nuclear EDM contribution.

We also remark that the last four of the molecules above have ${}^3\Delta_1$ as their ground or metastable state and thus have doublets of opposite parities and very small energy gaps (which may be manipulated by external electric and magnetic fields to scan for resonance with the axionic dark matter field). Accordingly, if the axion mass $\epsilon$ is of the order of these doublet splittings, the coefficient $2/3$ in the results\ \eqref{D_mol_EDM}--\eqref{res} should be replaced by $1/2$ and the first rotational energy $E_{\rm rot}$ by the energy $E_{\rm dbt}$ of the ${}^3\Delta_1$ doublet splitting. The value of $E_{\rm dbt}$ for HfF+ is given in Ref.\ \cite{HfFdoublet}, that for ThF+ in Refs.\ \cite{ThFdoublet_old,ThFdoublet}, for ThO in Refs.\ \cite{ThOdoublet_old,ThOdoublet} and for WC in Ref\ \cite{WCdoublet}.

The values $\bar{X}_{\rm HF}\approx1.7\,a_B$, $\bar{X}_{\rm LiF}\approx2.9\,a_B$, $\bar{X}_{\rm TlF}\approx3.9\,a_B$, $\bar{X}_{\rm ThO}\approx3.5\,a_B$, $\bar{X}_{\rm YbF}\approx3.8\,a_B$ and $\bar{X}_{\rm BaF}\approx4.1\,a_B$ ($a_B$ is the Bohr radius) are taken from the NIST database\ \cite{NIST}. The values $\bar{X}_{\rm HfF^+}\approx3.4\,a_B$, $\bar{X}_{\rm ThF^+}\approx3.8\,a_B$ and $\bar{X}_{\rm WC}\approx3.2\,a_B$ are taken from Refs.\ \cite{HfF+},\ \cite{ThF+} and\ \cite{WC}, respectively.

The value $\bar{d}_{\rm HF}\approx0.7\,ea_B$ is taken from Ref.\ \cite{HF}, the value $\bar{d}_{\rm LiF}\approx2.5\,ea_B$ from Ref.\ \cite{LiFd}, the value $\bar{d}_{\rm TlF}\approx3.0\,ea_B$ from Refs.\ \cite{BarrettMandel,Fitzky}, the value $\bar{d}_{\rm HfF{}^+}\approx1.4\,ea_B$ from Ref.\ \cite{Cornell}, the value $\bar{d}_{\rm BaF}\approx1.3\,ea_B$ from Ref.\ \cite{BaFdip}, the value $\bar{d}_{\rm ThO}\approx1.1\,ea_B$ from Ref.\ \cite{ThO}, the value $\bar{d}_{\rm WC}\approx1.6\,ea_B$ from Ref.\ \cite{WCdip}. For the molecules ThF${}^+$ and YbF, we assume the generic value $\bar{d}_{\rm ThF{}^+, YbF}\sim2\,ea_B$.

 The values for the Schiff moment $S_{\rm Tl}$ and interaction strength $W_S$ for TlF are taken from Refs.\ \cite{SFK,Cov}. The value of $W_S$ for LiF may be estimated by scaling with the nuclear charge $Z$ using the formula in Ref.\ \cite{SFK}. The EDM of Li may be estimated using formula\ \eqref{d}: $d_{\rm Li} \approx 3\times 10^{-16}\theta\, e\,{\rm cm}$. From Eq.\ \eqref{Schiff}, we obtain an estimate $S_{\rm F} \sim 3\times 10^{-4}\theta\,e\,{\rm fm}^3$. 
 
In Figs.\ \ref{LiF_graph} and\ \ref{TlF_graph} below, we show the behavior of $D^{\rm EDM}_{\rm mol}/d_1$ and $D^{\rm EDM}_{\rm mol}/D^{\rm SCHIFF}_{\rm mol}$ in LiF and TlF. The pictures for the other molecules will be similar to that of LiF. The quantities of interest, i.e., the large $\epsilon$ asymptotic value and resonance value of $D^{\rm EDM}_{\rm mol}/d_1$ and the position of the resonances (rotation or doublet), are summarized in Table.\ \ref{Results table}. Note that we have assumed that $E_{{\rm rot},{\rm dbt}}/\Gamma \approx 10^6$ (trapped molecule, $\Gamma$ is due to axion velocity distribution).


\begin{figure}[htb]
    \centering
    \includegraphics[scale=0.45]{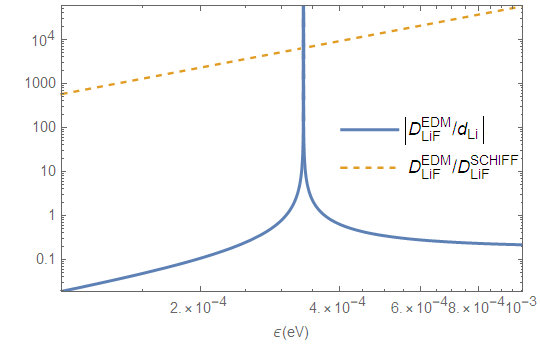}
    \caption{Ratios of molecular EDM induced by nuclear EDM with the nuclear EDM and with the molecular EDM induced by nuclear Schiff moment in LiF.}
    \label{LiF_graph}
\end{figure}
\begin{figure}[htb]
    \centering
    \includegraphics[scale=0.35]{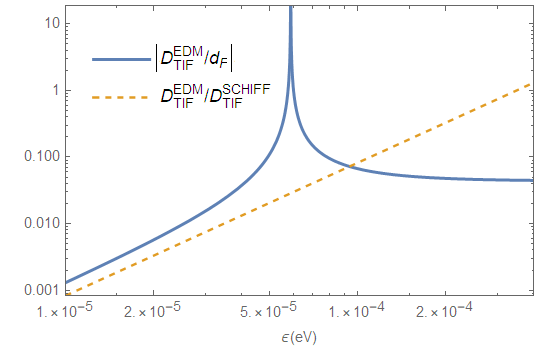}
    \caption{Ratios of molecular EDM induced by nuclear EDM with the nuclear EDM and with the molecular EDM induced by nuclear Schiff moment in TlF.}
    \label{TlF_graph}
\end{figure}
\begin{table}
\begin{tabular}{|c|P{2.2cm}|P{1.5cm}|P{1.5cm}|}
\hline
   & Resonance position (eV) & Large $\omega$ value & Resonance value\\
   \hline
HF (${}^1\Sigma^+$) & $5.2\times 10^{-3}$ & 0.8 & $8\times 10^{5}$ \\
\hline
LiF (${}^1\Sigma^+$) & $3.4\times 10^{-4}$ & 0.2 & $2\times 10^{5}$ \\
\hline
YbF (${}^2\Sigma^+$) & $6.0\times 10^{-5}$ & 0.04 & $4\times 10^{4}$ \\
\hline
BaF (${}^2\Sigma^+$) & $5.3\times 10^{-5}$ & 0.02 & $2\times 10^{4}$ \\
\hline
TlF (${}^1\Sigma^+$) & $5.6\times 10^{-5}$ & 0.06 & $6\times 10^{4}$ \\
\hline
HfF${}^{+}$ (${}^1\Sigma^+$) & $7.5\times 10^{-5}$ & 0.04 & $4\times 10^{4}$ \\
\hline
HfF${}^{+}$ (${}^3\Delta_1$) & $4.1\times 10^{-11}$ & 0.06 & $6\times 10^{4}$ \\
\hline
ThF${}^{+}$ (${}^1\Sigma^+$) & $5.8\times 10^{-5}$ & 0.04 & $4\times 10^{4}$ \\
\hline
ThF${}^{+}$ (${}^3\Delta_1$) & $2.9\times 10^{-10}$ & 0.06 & $6\times 10^{4}$ \\
\hline
ThO (${}^1\Sigma^+$) & $7.6\times 10^{-5}$ & 0.03 & $3\times 10^{4}$ \\
\hline
ThO (${}^3\Delta_1$) & $7.7\times 10^{-10}$ & 0.05 & $5\times 10^{4}$ \\
\hline
WC (${}^3\Delta_1$) & $4.1\times 10^{-12}$ & 0.08 & $8\times 10^{4}$ \\
\hline
\end{tabular}
\caption{Position of the resonance(rotational or $\Omega$-doublet), large axion mass asymptotic and resonance values of the ratio $\left|D_{\rm mol}^{EDM}/d_1\right|$ between the magnitude of the molecular EDM induced by the oscillating nuclear EDM ${\bf d}_1$ and $d_1$ in several molecules.}\label{Results table}
\end{table}

We also note that the estimates presented in this paper may be readily extended to the case of polyatomic molecules (see, for example, Ref.\ \cite{TranFlambaum2019}). The advantage of polyatomic molecules is that since their spectra are very dense, the probability of a resonance with the axionic dark matter field is higher. Solids also have rich spectra of low-energy excitations and effects of nuclear motion (similar to effect in molecules).




\textbf{Acknowledgment:} This work is supported by the Australian Research Council, Gutenberg Fellowship and New Zealand Institute for Advanced Study. We thank Oleg Sushkov and Igor Samsonov for helpful discussions.

\section*{Appendix A}

In this appendix, we provide the derivation for the results\ \eqref{D_mol_EDM},\ \eqref{res} and\ \eqref{D_SCHIFF}.

The total Hamiltonian of a diatomic molecule is given by
\begin{equation}
\begin{aligned}
  H&=\frac{\mathbf{P}_{1}^{2}}{2{{M}_{1}}}+\frac{\mathbf{P}_{2}^{2}}{2{{M}_{2}}}+\sum\limits_{i=1}^{{{N}_{e}}}{\frac{\mathbf{p}_{i}^{2}}{2{{m}_{e}}}}\\
  &+{{V}_{0}}+{{V}_{\text{EDM}}}+{{V}_{\text{SCHIFF}}} \,,\\ 
  {{V}_{0}}&=-\sum\limits_{i=1}^{{{N}_{e}}}{\frac{{{Z}_{1}}{{e}^{2}}}{\left| {{\mathbf{r}}_{i}}-{{\mathbf{R}}_{2}} \right|}}-\sum\limits_{i=1}^{{{N}_{e}}}{\frac{{{Z}_{2}}{{e}^{2}}}{\left| {{\mathbf{r}}_{i}}-{{\mathbf{R}}_{2}} \right|}}\\
  &+\sum\limits_{i>j}^{{{N}_{e}}}{\frac{{{e}^{2}}}{\left| {{\mathbf{r}}_{i}}-{{\mathbf{r}}_{j}} \right|}}+\frac{{{Z}_{2}}{{Z}_{1}}{{e}^{2}}}{\left| {{\mathbf{R}}_{2}}-{{\mathbf{R}}_{1}} \right|} \,,\\ 
  {{V}_{\text{EDM}}}&=\frac{{{\mathbf{d}}_{1}}\cdot {{\nabla }_{{{\mathbf{R}}_{1}}}}{{V}_{0}}}{{{Z}_{1}}e}+\frac{{{\mathbf{d}}_{2}}\cdot {{\nabla }_{{{\mathbf{R}}_{2}}}}{{V}_{0}}}{{{Z}_{2}}e}\,, \\ 
  {{V}_{\text{SCHIFF}}}&={{W}_{S}}\mathbf{S}\cdot \frac{{{\mathbf{R}}_{1}}-{{\mathbf{R}}_{2}}}{\left| {{\mathbf{R}}_{1}}-{{\mathbf{R}}_{2}} \right|}\,, \\ 
\end{aligned}
\end{equation}
where the nuclear positions ${\bf R}_{1,2}$, nuclear momenta ${\bf P}_{1,2}$, electrons position ${\bf r}_i$ and electron momenta ${\bf p}_i$ are defined in the laboratory frame.

A change of coordinates to the center-of-mass frame as described in Ref.\ \cite{TranFlambaum2019}, gives, after discarding the free motion of the molecule
\begin{equation}
\begin{aligned}
   H&={{H}_{0}}+{{V}_{\text{EDM}}}+{{V}_{\text{Schiff}}} \,,\\ 
  {{H}_{0}}&=\frac{{{\mathbf{Q}}^{2}}}{2{{\mu }_{N}}}+\sum\limits_{i=1}^{{{N}_{e}}}{\frac{\mathbf{q}_{i}^{2}}{2{{\mu }_{e}}}}+\sum\limits_{i\ne j}^{{{N}_{e}}}{\frac{{{\mathbf{q}}_{i}}{{\mathbf{q}}_{j}}}{{{M}_{N}}}+{{V}_{0}}} \,,\\ 
\end{aligned}
\end{equation}
where $V_0$, $V_{\rm EDM}$ and $V_{\rm SCHIFF}$ are now functions of the new variables ${\bf X}= {\bf R}_1-{\bf R}_2$ and ${\bf x}_i={\bf r}_i-\left(M_1{\bf R}_1+M_2{\bf R}_2\right)/M_N$. The momenta $\bf Q$ and ${\bf q}_i$ are conjugate to ${\bf X}$ and ${\bf x}_i$, respectively. For convenience, we have defined $M_N=M_1+M_2$ and $M_T=M_N+N_em_e$.

The EDM induced by ${\bf d}_1$ and ${\bf d}_2$ is given by
\begin{equation}\label{D_ind}
{\bf D}_{\rm ind}^{\rm EDM}=  2\sum_n \frac{E_{0n}\bra{0}V_{\rm EDM}\ket{n}\bra{n}{\bf d}\ket{0}}{E_{0n}^2 -\epsilon^2}\,,
\end{equation}
where
\begin{equation}\label{Def_d}
\mathbf{d}=-{{\zeta }_{e}}\sum\limits_{i=1}^{{{N}_{e}}}{{{\mathbf{x}}_{i}}}+{{\zeta }_{N}}\mathbf{X}
\end{equation}
is the molecule's total EDM operator. Here, ${{\zeta }_{e}}=e\left( {{M}_{N}}+{{Z}_{N}}{{m}_{e}} \right)/{{M}_{T}}$ and ${{\zeta }_{N}}=e\left( {{M}_{2}}{{Z}_{1}}-{{M}_{1}}{{Z}_{2}} \right)/{{M}_{N}}$.

Using the relations (the terms proportional to the molecule's total momentum have been discarded)
\begin{equation}
{{\mathbf{P}}_{I}}=-{{\left( -1 \right)}^{I}}\mathbf{Q}-\frac{{{M}_{I}}}{{{M}_{N}}}\sum\limits_{i=1}^{{{N}_{e}}}{{{\mathbf{q}}_{i}}}\,,
\end{equation}
we may write
\begin{equation}
\begin{aligned}\label{V_EDM}
   {{V}_{\text{EDM}}}&=\frac{{{\mathbf{d}}_{1}}\cdot {{\nabla }_{{{\mathbf{R}}_{1}}}}{{V}_{0}}}{{{Z}_{1}}e}+\frac{{{\mathbf{d}}_{2}}\cdot {{\nabla }_{{{\mathbf{R}}_{2}}}}{{V}_{0}}}{{{Z}_{2}}e}\\
   &=\frac{i{{\mathbf{d}}_{1}}\cdot \left[ {{\mathbf{P}}_{1}},{{H}_{0}} \right]}{{{Z}_{1}}e\hbar }+\frac{i{{\mathbf{d}}_{2}}\cdot \left[ {{\mathbf{P}}_{2}},{{H}_{0}} \right]}{{{Z}_{2}}e\hbar } \\ 
 & =\frac{i{{\mathbf{d}}_{1}}\cdot \left[ \mathbf{Q}-\frac{{{M}_{1}}}{{{M}_{N}}}\sum\limits_{i=1}^{{{N}_{e}}}{{{\mathbf{q}}_{i}}},{{H}_{0}} \right]}{{{Z}_{1}}e\hbar }\\
 &-\frac{i{{\mathbf{d}}_{2}}\cdot \left[ \mathbf{Q}+\frac{{{M}_{2}}}{{{M}_{N}}}\sum\limits_{i=1}^{{{N}_{e}}}{{{\mathbf{q}}_{i}}},{{H}_{0}} \right]}{{{Z}_{2}}e\hbar }\,. \\ 
\end{aligned}
\end{equation}

Substituting formula\ \eqref{V_EDM} into Eq.\ \eqref{D_ind}, we obtain
\begin{equation}\label{22}
\begin{aligned}
  \mathbf{D}_{\text{ind}}^{\text{EDM}}&=\frac{2}{e\hbar }\sum\limits_{n}{\text{Im}\left( \left\langle  0 \right|\Pi \left| n \right\rangle \left\langle  n \right|\mathbf{d}\left| 0 \right\rangle  \right)} \\ 
  &+\frac{2}{e\hbar }\sum\limits_{n}{\frac{{{\epsilon }^{2}}}{E_{0n}^{2}-{{\epsilon }^{2}}}\text{Im}\left( \left\langle  0 \right|\Pi \left| n \right\rangle \left\langle  n \right|\mathbf{d}\left| 0 \right\rangle  \right)}\,, \\ 
\end{aligned}
\end{equation}
where we have defined
\begin{equation}
\begin{aligned}
\Pi &=\frac{{{\mathbf{d}}_{1}}}{{{Z}_{1}}}\cdot \left( \mathbf{Q}-\frac{{{M}_{1}}}{{{M}_{N}}}\sum\limits_{i=1}^{{{N}_{e}}}{{{\mathbf{q}}_{i}}} \right)\\
&-\frac{{{\mathbf{d}}_{2}}}{{{Z}_{2}}}\cdot \left( \mathbf{Q}+\frac{{{M}_{2}}}{{{M}_{N}}}\sum\limits_{i=1}^{{{N}_{e}}}{{{\mathbf{q}}_{i}}} \right)\,.
\end{aligned}
\end{equation}

The $\epsilon$-independent term in Eq.\ \eqref{22} may be written as
\begin{equation}
\begin{aligned}
  &\frac{2}{e\hbar }\sum\limits_{n}{\text{Im}\left( \left\langle  0 \right|\Pi \left| n \right\rangle \left\langle  n \right|\mathbf{d}\left| 0 \right\rangle  \right)} \\ 
  &=-\frac{i}{e\hbar }\frac{{{\mathbf{d}}_{1}}}{{{Z}_{1}}}\cdot \left\langle  0 \right|\left[ \mathbf{Q}-\frac{{{M}_{1}}}{{{M}_{N}}}\sum\limits_{i=1}^{{{N}_{e}}}{{{\mathbf{q}}_{i}}},\mathbf{d} \right]\left| 0 \right\rangle  \\ 
  &+\frac{i}{e\hbar }\frac{{{\mathbf{d}}_{2}}}{{{Z}_{2}}}\cdot \left\langle  0 \right|\left[ \mathbf{Q}+\frac{{{M}_{2}}}{{{M}_{N}}}\sum\limits_{i=1}^{{{N}_{e}}}{{{\mathbf{q}}_{i}}},\mathbf{d} \right]\left| 0 \right\rangle  \\ 
  &=-\left( 1+\frac{{{M}_{1}}{{Z}_{T}}}{{{M}_{T}}{{Z}_{1}}} \right){{\mathbf{d}}_{1}}-\left( 1+\frac{{{M}_{2}}{{Z}_{T}}}{{{M}_{T}}{{Z}_{2}}} \right){{\mathbf{d}}_{2}} \\ 
\end{aligned}\,,
\end{equation}
where $Z_T=Z_1+Z_2-N_e$. For neutral molecule ($Z_T=0$), this term exactly cancels the contribution of ${\bf d}_1$ and ${\bf d}_2$ to the total molecular EDM.

Using the relations
\begin{equation}
\begin{aligned}
\mathbf{Q}&+\left( -{{1}^{I}} \right)\frac{{{M}_{I}}}{{{M}_{N}}}\sum\limits_{i=1}^{{{N}_{e}}}{{{\mathbf{q}}_{i}}}\\
&=\frac{i}{\hbar }\left[ {{H}_{0}},{{\mu }_{N}}\mathbf{X}+\left( -{{1}^{I}} \right)\frac{{{M}_{I}}{{\mu }_{e}}}{{{M}_{T}}}\sum\limits_{i=1}^{{{N}_{e}}}{{{\mathbf{x}}_{i}}} \right]
\end{aligned}
\end{equation}
and the definition\ \eqref{Def_d} (which may be used to express $\sum\limits_{i=1}^{{{N}_{e}}}{{{\mathbf{x}}_{i}}}$ in terms of $\bf d$ and $\bf X$), we obtain
\begin{equation}
\left\langle  0 \right|\Pi \left| n \right\rangle =\frac{i{{E}_{0n}}{{\mu }_{N}}}{e\hbar\sqrt{Z_1Z_2} }\left\langle  0 \right|{\boldsymbol{\delta }\cdot \mathbf{X}}+\frac{{{m}_{e}}}{\sqrt{{{M}_{1}}{{M}_{2}}}}\mathbf{\Delta }\cdot \mathbf{d}\left| n \right\rangle\,,
\end{equation}
where
\begin{equation}
\boldsymbol{\delta }\approx \frac{{{Z}_{2}}{{\mathbf{d}}_{1}}-{{Z}_{1}}{{\mathbf{d}}_{2}}}{\sqrt{{{Z}_{1}}{{Z}_{2}}}}\,,
\end{equation}
and
\begin{equation}
\mathbf{\Delta }\approx \frac{{{Z}_{2}}{{M}_{1}}{{\mathbf{d}}_{1}}+{{Z}_{1}}{{M}_{2}}{{\mathbf{d}}_{2}}}{\sqrt{{{M}_{1}}{{M}_{2}}{{Z}_{1}}{{Z}_{2}}}}\,.
\end{equation}

As a result, the $\epsilon$-dependent term in Eq.\ \eqref{22} may be written as
\begin{equation}
\begin{aligned}
  & \frac{2}{e\hbar }\sum\limits_{n}{\frac{{{\epsilon }^{2}}\text{Im}\left( \left\langle  0 \right|\Pi \left| n \right\rangle \left\langle  n \right|\mathbf{d}\left| 0 \right\rangle  \right)}{{E_{0n}^{2}}-{{\epsilon }^{2}}}} \\ 
 & \approx -\frac{{{\epsilon }^{2}}{{m}_{e}}{{\mu }_{N}}}{{{e}^{2}}{{\hbar }^{2}}\sqrt{{{M}_{1}}{{M}_{2}}{{Z}_{1}}{{Z}_{2}}}}\left( \alpha\mathbf{\Delta }+\beta\boldsymbol{\delta } \right) \,,\\ 
\end{aligned}
\end{equation}
where
\begin{equation}
\alpha=2\sum\limits_{n}{\frac{{{E}_{n0}}\left\langle  0 \right|\mathbf{d}\left| n \right\rangle \left\langle  n \right|\mathbf{d}\left| 0 \right\rangle }{E_{n0}^{2}-{{\epsilon }^{2}}}}
\end{equation}
is the molecular polarizability tensor and 
\begin{equation}
\beta=\frac{2\sqrt{M_1M_2}}{{{m}_{e}}}\sum\limits_{n}{\frac{{{E}_{n0}} \left\langle  0 \right|e\mathbf{X}\left| n \right\rangle \left\langle  n \right|\mathbf{d}\left| 0 \right\rangle  }{E_{n0}^{2}-{{\epsilon }^{2}}}}\,.
\end{equation}

If $\epsilon\ll 1{\rm eV}$ then $\beta\boldsymbol{\delta}$ dominates over $\alpha\mathbf{\Delta}$ because of the factor $\sqrt{M_1M_2}/m_e$. Approximating the sum over states $\beta$ by the term involved the first rotational state and using the Born-Oppenheimer wavefunction, we obtain the result\ \eqref{D_mol_EDM}.

If the oscillation of the nuclear EDMs is in resonance with the first rotational energy, $\epsilon = E_{\rm rot}$, then, following Refs.\ \cite{FlambaumSamsonov2018,TranFlambaum2019}, the formula\ \eqref{D_ind} is replaced (for a neutral molecule) by the following relation between the oscillation amplitude of $\mathbf{D}_{\text{mol}}^{\text{EDM}}$ and $\bf d$
\begin{equation}
D_{\text{mol}}^{\text{EDM}}=\frac{2}{{{\Gamma }}}\left|\left\langle  0 \right|\mathbf{d}\left| 1 \right\rangle \left\langle  1 \right|{{V}_{\text{EDM}}}\left| 0 \right\rangle  \right|\,,
\end{equation}
where the ket $\ket{1}$ denotes the first rotational state and ${\Gamma }$ is its width. Note that if ${\Gamma }$ is the natural width and ${\bf d}_{1,2}$ have time dependence $\cos\omega t$ then $\mathbf{D}_{\text{mol}}^{\text{EDM}}$ is proportional to $\sin\omega t$. Carrying out the same analysis as above, we obtain the result\ \eqref{res}.

Finally, we may estimate the contribution to the molecular EDM of the oscillating Schiff moment as
\begin{equation}
\begin{aligned}
  \mathbf{D}_{\text{mol}}^{\text{SCHIFF}}&=2\sum\limits_{n}{\frac{{{E}_{0n}}\operatorname{Re}\left( \left\langle  0 \right|{{V}_{\text{SCHIFF}}}\left| n \right\rangle \left\langle  n \right|\mathbf{d}\left| 0 \right\rangle  \right)}{E_{0n}^{2}-{{\epsilon }^{2}}}} \\ 
  &=2{{W}_{S}}\mathbf{S}\sum\limits_{n}{\frac{{{E}_{0n}}\left\langle  0 \right|\mathbf{\hat{X}}\left| n \right\rangle \left\langle  n \right|\mathbf{d}\left| 0 \right\rangle}{E_{0n}^{2}-{{\epsilon }^{2}}}} \\ 
  &\approx \frac{2\bar{d}{{W}_{S}}\mathbf{S}}{3}\frac{{{E}_{\text{rot}}}}{E_{\text{rot}}^{2}-{{\epsilon }^{2}}}\,, \\ 
\end{aligned}
\end{equation}
where $\hat{\bf X}$ is the unit vector along the inter-nuclear axis. Note that we have taken into account only the contribution of the first rotational state. In the case where $\epsilon = E_{\rm rot}$, we need to replace the the factor $\frac{{{E}_{\text{rot}}}}{E_{\text{rot}}^{2}-{{\epsilon }^{2}}}$ by $\Gamma^{-1}$.

\end{document}